\documentclass[letter,twocolumn]{jpsj3}

\usepackage{amsmath}
\usepackage{amsfonts}
\usepackage{amssymb}
\usepackage{txfonts}
\usepackage{bm}
\usepackage{tabularx}
\usepackage{graphicx,color}

\def\Vec#1{\bm{#1}}
\def\Hc2{H_\mathrm{c2}}
\def\Tc{T_\mathrm{c}}

\author{
	Shunichiro Kittaka$^{1,\thanks{E-mail: kittaka@issp.u-tokyo.ac.jp}}$, 
	Yuya Aoki$^{1}$,
	Naoki Kase$^{1}$,
	Toshiro Sakakibara$^{1}$, 
	Taku Saito$^{2}$, 
	Hideto Fukazawa$^{2,3}$, \\
	Yoh Kohori$^{2,3}$, 
	Kunihiro Kihou$^{3,4}$, 
	Chul-Ho Lee$^{3,4}$, 
	Akira Iyo$^{3,4}$,
	Hiroshi Eisaki$^{3,4}$, \\
	Kazuhiko Deguchi$^5$, 
	Noriaki K. Sato$^5$, 
	Yasumasa Tsutsumi$^6$, and 
	Kazushige Machida$^7$ 
}

\inst{$^{1}$Institute for Solid State Physics, University of Tokyo, Kashiwa, Chiba 277-8581, Japan\\
      $^{2}$Department of Physics, Chiba University, Chiba 263-8522, Japan \\
      $^{3}$JST, Transformative Research-Project on Iron Pnictides (TRIP), Chiyoda, Tokyo 102-0075, Japan \\
      $^{4}$National Institute of Advanced Industrial Science and Technology, Tsukuba, Ibaraki 305-8568, Japan\\
      $^{5}$Department of Physics, Graduate School of Science, Nagoya University, Nagoya 464-8602, Japan\\
      $^{6}$Condensed Matter Theory Laboratory, RIKEN, Wako, Saitama 351-0198, Japan\\
      $^{7}$Department of Physics, Okayama University, Okayama 700-8530, Japan
}
\begin{document}

\title{Thermodynamic Study of Nodal Structure and Multiband Superconductivity of KFe$_2$As$_2$}

\date{\today}

\abst{
The temperature, field, and field-orientation dependences of the electronic specific heat $C_{\rm e}$ of the iron-pnictide superconductor KFe$_2$As$_2$ have been investigated.
Thermodynamic evidence of the presence of line nodes is obtained from the $T$ and $\sqrt{H}$ linear dependences of $C_{\rm e}/T$ in the low-$T$ and low-$H$ region. 
Under a magnetic field rotated within the tetragonal $ab$ plane, a fourfold oscillation is observed in $C_{\rm e}$ with a sign change at $0.08\Tc$.
On the basis of the Doppler-shift analysis, 
the observed $C_{\rm e}$ minima in $H \parallel [100]$ at low $T$ indicate the presence of line nodes somewhere on the Fermi surface 
where the Fermi velocity is parallel to the $[100]$ direction;
this is consistent with the octet-line-node scenario proposed recently by a photoemission experiment.
In addition, the low-$T$ $C_{\rm e}/T$ exhibits an unusual upturn on cooling at moderate fields only for $H \parallel ab$,
which is understood in terms of the strong Pauli paramagnetic effect on multiband superconductivity.
}

\maketitle

The discovery of superconductivity at 26~K in LaFeAsO$_{1-x}$F$_x$ \cite{Kamihara2008JACS} has triggered much interest in 
the study of high-temperature superconductors consisting of iron pnictides. 
The (Ba$_{1-x}$K$_x$)Fe$_2$As$_2$ series has attracted much attention because of a striking change in its superconducting gap structure upon varying the dopant $x$.\cite{Rotter2008PRL} 
At approximately $x \sim 0.4$, $\Tc$ reaches as high as 38~K and the superconducting gap is suggested to be nodeless~\cite{Popovich2010PRL,Li2011PRB}. 
By contrast, the end member KFe$_2$As$_2$ exhibits superconductivity at a relatively low temperature $T$ below 3.4~K, and 
it has been indicated to have line nodes in the gap~\cite{Fukazawa2009JPSJ,Hashimoto2010PRB}.
In this series, the strong Pauli paramagnetic effect occurs for fields applied to the FeAs plane.\cite{Terashima2013PRB2,Terashima2009JPSJ,Zocco2013PRL,Burger2013PRB}

To resolve the pairing mechanism, the location of gap nodes on the Fermi surface provides an important clue.
By the de Haas van Alphen (dHvA) oscillation and angle-resolved photoemission-spectroscopy (ARPES) experiments, \cite{Sato2009PRL,Okazaki2012Science, Terashima2013PRB}
KFe$_2$As$_2$ has been revealed to have three cylinders at the $\Gamma$ point, which are labelled as $\alpha$ (inner), $\zeta$ (middle), and $\beta$ (outer) bands, 
and four small cylinders near the $X$ point ($\epsilon$ band), 
for which Sommerfeld coefficients $\gamma$ are estimated to be 9.1 ($\alpha$), 19.2 ($\zeta$), 27.6 ($\beta$), and $9.6 \times 4$ ($\epsilon$) mJ/(mol~K$^2$) 
from dHvA measurements.\cite{Terashima2013PRB}
Such a Fermi-surface topology induces the presence of multiple superconducting gaps.
Recently, Okazaki \textit{et al}. have determined the gap structure on the $\alpha$, $\zeta$, and $\beta$ bands by the laser ARPES experiment~\cite{Okazaki2012Science}.
They suggested a nodeless gap on the $\alpha$ and $\beta$ bands, and a nodal gap with octet line nodes on the $\zeta$ band 
(the gap structure on the $\epsilon$ band was not investigated).
These results led them to conclude that the gap symmetry of KFe$_2$As$_2$ is of the $s$-wave type with accidental line nodes,
although the possibility of the $d$-wave state has also been proposed.\cite{Reid2012PRL}

To obtain thermodynamic evidence of the gap structure of KFe$_2$As$_2$,
we performed a field-angle-resolved specific-heat measurement $C_{\rm e}(T, H, \phi)$,\cite{Sakakibara2007JPSJ} 
which detects the quasiparticle (QP) density of states (DOS) at the Fermi level.
We revealed the $T$ and $\sqrt{H}$ linear dependences of $C_{\rm e}/T$ at low temperatures and low fields,
proving the existence of line nodes in the gap.
In addition, under a rotating magnetic field within the $ab$ plane, 
we detected a fourfold $C_{\rm e}(\phi)$ oscillation that is minimum for fields along the $[100]$ direction at low $T$,
suggesting the presence of line nodes somewhere in the gap where the Fermi velocity $\Vec{v}_{\rm F} \parallel [100]$, 
on the basis of the Doppler-shift analysis.
A sign change of the oscillation observed at $T \sim 0.08\Tc$ demonstrates that 
the Doppler-shift effect is indeed the major factor behind the field-angle dependence of $C_{\rm e}(\phi)$ in the low-$T$ region.
Moreover, at moderate fields for $H \parallel ab$, we found an unusual upturn in the $T$ variation of $C_{\rm e}/T$ upon cooling, 
which is explained in the framework of multiband superconductivity under a strong Pauli paramagnetic effect on a minor gap.

Single crystals of KFe$_2$As$_2$ were grown by the self-flux method.
Three samples, A (6.6~mg), B (10.2~mg), and C (8.0~mg), were used in the present study.
All the samples have the shape of a flat slab with the shortest dimension along the $c$-axis, and 
the directions of the crystal axes have been determined by X-ray single-crystal diffraction analysis.
The specific heat was measured by the relaxation and quasi-adiabatic heat-pulse methods in a dilution refrigerator under 
magnetic fields generated by a vector magnet.
In all the data presented below, the nuclear contribution calculated using a nuclear spin Hamiltonian, $C_{\rm n}=(1.4+2.66H^2)/T^2$~$\muup$J/(mol K), and 
the addenda specific heat were subtracted.

\begin{figure}[t]
\includegraphics[width=3.2in]{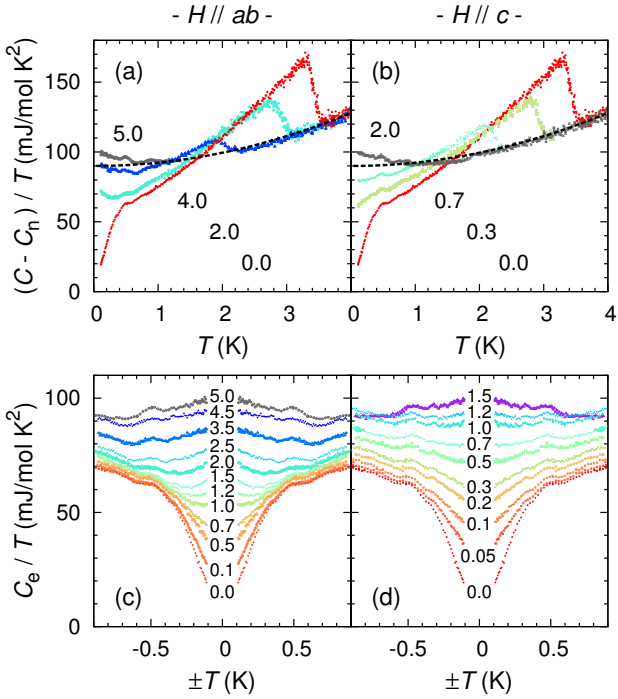}
\caption{
(Color online) 
Temperature dependences of $(C-C_{\rm n})/T$ under magnetic fields applied parallel to the (a) $[110]$ and (b) $[001]$ directions. Dashed lines represent $C/T = \beta_1 T^2 + \gamma$.
Low-temperature parts of $(C-C_{\rm n}-\beta_1 T^3)/T$ (=$C_{\mathrm e}/T$) plotted as functions of $T$ and $-T$ for (c) $H \parallel ab$ and (d) $H \parallel c$.
Numbers labeling the curves represent the applied magnetic field in tesla.
These data were obtained using sample~A.
}
\label{Tdep}
\end{figure}

Figures 1(a) and 1(b) show the temperature dependences of $(C-C_{\rm n})/T$ measured using sample~A at several magnetic fields for $H \parallel ab$ and $H \parallel c$, respectively.
The phonon contribution is represented by the dashed line, which is estimated by fitting the normal-state data at 2~T for $H \parallel c$ 
in the range 2~K $\le T \le 4$~K using the function $C/T = \beta_1 T^2 + \gamma$.
Here, $\beta_1$ is the coefficient of the Debye term. 
The Debye temperature and the electronic specific-heat coefficient $\gamma$ are estimated to be 160~K and 90~mJ/(mol~K$^2$), respectively.
At low $T$ below 0.6~K, the normal-state $(C-C_{\rm n})/T$ increases slightly on cooling.
This might be related to the non-Fermi-liquid behavior of KFe$_2$As$_2$~\cite{Dong2010PRL}.

At zero field, a sharp superconducting transition is observed at $\Tc=3.4$~K,
indicating the high quality of the present sample. 
The specific-heat jump $\Delta C/\gamma \Tc$ is estimated to be 0.6, which is much smaller than the BCS expectation ($=1.43$).
With decreasing $T$, $(C-C_{\rm n})/T$ exhibits a rapid decrease below 0.5~K, 
which, along with the small specific-heat jump, is a hallmark of the presence of a minor gap. 
These results are in good agreement with previous reports~\cite{Fukazawa2009JPSJ,Hardy2013PRL}.
The $T$-linear behavior at low $T$, whose extrapolation to 0~K gives a finite residual term, implies the presence of line nodes in the gap.

Figures 1(c) and 1(d) show the low-$T$ parts of $C_{\mathrm e}/T=(C-C_{\rm n}-\beta_1 T^3)/T$ at various magnetic fields applied perpendicular and parallel to the $c$-axis, respectively.
Here, the data are plotted as functions of $\pm T$. 
The upper critical fields $\Hc2$ are about 5~T for $H \parallel ab$ and 1.4~T for $H \parallel c$.
Note that Figs. 1(c) and 1(d) correspond to the energy spectra of spatially averaged QP DOS, $N(E)$, at low energy
because the $T$ dependence of $C_{\rm e}/T$ at low $T$ can be regarded as $N(E)$ at low energy.\cite{Kittaka2013arXiv}
The small V-shaped structure in $C_{\rm e}/T$ for $|T| \le 0.5$~K, attributed to the contribution from minor gaps, is significantly suppressed by a relatively low field ($\sim 0.2\Hc2$) in both field directions.
However, in the intermediate-field region, 
a striking peak, instead of a V-shaped dip, appears in $C_{\mathrm e}/T$ at $T \sim 0$ for $H \parallel ab$, 
though it is absent for $H \parallel c$. 
According to the relation $C_{\rm e}/T \propto N(E)$ at low energy, 
this result indicates that QP DOS at low energy is enhanced by magnetic fields for $H \parallel ab$.

\begin{figure}
\includegraphics[width=3.2in]{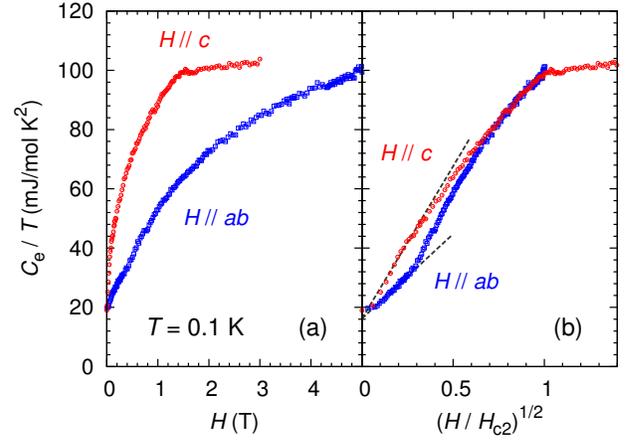}
\caption{
(Color online) (a) Field dependences of the specific heat of sample~A for $H \parallel [110]$ and $H \parallel [001]$ at 0.1~K.
(b) Same data plotted as a function of $\sqrt{H/\Hc2}$. 
Dashed lines represent the function $a\sqrt{H}+b$.
}
\label{Hdep}
\end{figure}

Figure 2(a) shows the field dependences of $C_{\mathrm e}/T$ at 0.1~K for $H \parallel ab$ and $H \parallel c$.
For both field orientations, $C_{\mathrm e}/T$ exhibits a rapid increase at low fields; 
it is nearly proportional to the square root of $H$ [dashed lines in Fig.~2(b)].
The observed $\sqrt{H}$ behavior 
demonstrates the occurrence of low-energy QP excitations around nodes.\cite{Volovik1993JETPL}
From the specific-heat measurement, the transition at $\Hc2$ seems to be of second order in both field directions,
though the first-order transition has been reported from magnetostriction measurements for $H \parallel ab$.\cite{Zocco2013PRL}

For $H \parallel ab$, an unusual upward curvature is observed in $C_{\rm e}(H)$ at $\sqrt{H/\Hc2} \sim 0.3$ [Fig.~2(b)].
These unusual $T$ and $H$ dependences of $C_{\rm e}/T$ for $H \parallel ab$ are reminiscent of those for CeCu$_2$Si$_2$:\cite{Kittaka2013arXiv}
a striking upturn in the low-$T$ $C_{\mathrm e}/T$ on cooling at moderate fields,
indicating an enhancement of $N(E)$ at low energy under $H$,
and an upward kink in $C_{\rm e}(H)$ at low fields.
Whereas CeCu$_2$Si$_2$ shows these anomalies in any field direction, KFe$_2$As$_2$ does so only in $H \parallel ab$.
Note that the anomalies are seen for these materials in the field direction under which the Pauli paramagnetic effect is prominent;
it is known that 
the $\Hc2$ limit is absent (present) in $H \parallel c$~($\parallel ab$) for KFe$_2$As$_2$, and it is observable in any field direction for CeCu$_2$Si$_2$.
Indeed, such an enhancement of the low-energy $N(E)$ at moderate fields can occur if the strong paramagnetic effect is operative for a multiband superconductor,
as explained below.

\begin{figure}
\includegraphics[width=3.2in]{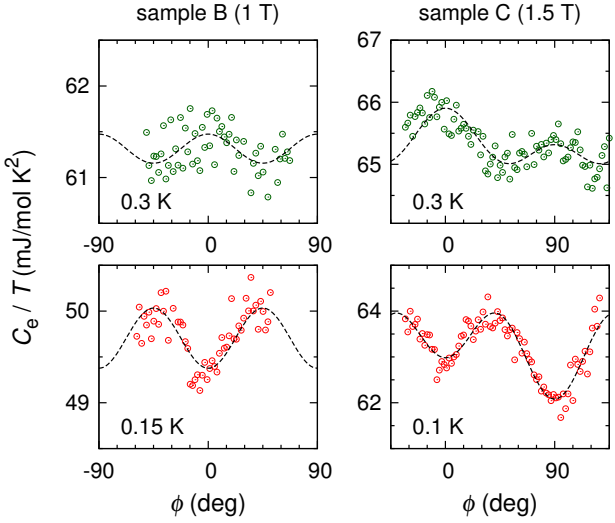}
\caption{
(Color online) 
Field-angle $\phi$ dependences  of $C_{\rm e}/T$ for samples B (left) and C (right). The angle $\phi$ is the in-plane field angle measured from the [100] axis.
Dashed lines are the fitting results (see text).
}
\label{phi}
\end{figure}

In general, in the vortex state, the $N(E)$ of a single-band superconductor with a gap size $\Delta$ has a V-shaped structure, 
i.e., $N(E) \propto |E|$ near $|E| \sim 0$ with an edge-singularity peak at $|E| \sim \Delta$.\cite{Nakai2006PRB}
With increasing field, this V-shaped structure widens and flattens, gradually approaching the normal-state DOS [$N(E) = {\rm constant}$].
In addition, the presence of a strong paramagnetic effect causes an energy shift of the edge peak toward $|E| \sim 0$.
In such a case, however, the enhancement of the low-energy $N(E)$ is not expected 
because the height of the edge peak is strikingly suppressed owing to a rapid destruction of the superconducting gap by the Zeeman effect.

For multiband superconductors, $N(E)$ is the superposition of multiple V-shaped structures with different $\Delta$'s, and 
they are also expected to widen and flatten with increasing $H$, similarly to the single-band case.
When a strong paramagnetic effect occurs for one of the minor gaps, 
the edge peak of the corresponding small V-shaped DOS would be shifted without a fatal suppression of its height;
the minor gap can remain largely opened by the assistance of the major gap, causing the DOS of the minor gap to be piled up around its edge peak. 
When the piled-up peak approaches $|E| \sim 0$ owing to the Zeeman effect with sufficient support from the major gap, 
the total DOS has the peak near $|E| \sim 0$ superposed on the large V-shaped structure of the major gap.
Thus, the enhancement of the low-energy $N(E)$ can be accomplished in the framework of multiband superconductivity with a strong paramagnetic effect.
An upward kink in $C_{\rm e}(H)$ may reflect the characteristic field at which the low-energy $N(E)$ starts to be enhanced by the Zeeman effect.

Let us turn our attention to the field-orientation dependence of $C_{\mathrm e}/T$ for KFe$_2$As$_2$.
Figure 3 shows $C_{\mathrm e}(\phi)/T$ in a rotating magnetic field within the $ab$ plane obtained using samples B and C 
measured in the ranges of $-60^\circ \lesssim \phi \lesssim 60^\circ$ and $-45^\circ \lesssim \phi \lesssim 135^\circ$, respectively. 
Here, $\phi$ is the azimuthal angle between the magnetic field and the [100] axis.
Both samples have a component of a fourfold oscillation in $C_{\rm e}(\phi)$.
Although a twofold contribution is observed in the $C_{\rm e}(\phi)$ of sample C, 
it can be attributed to the misalignment of the field direction with respect to the $ab$ plane. 
At low $T$ below 0.15~K, $C_{\rm e}(\phi)$ becomes minimum when the field is applied along the $[100]$ direction.

\begin{figure}
\includegraphics[width=3.2in]{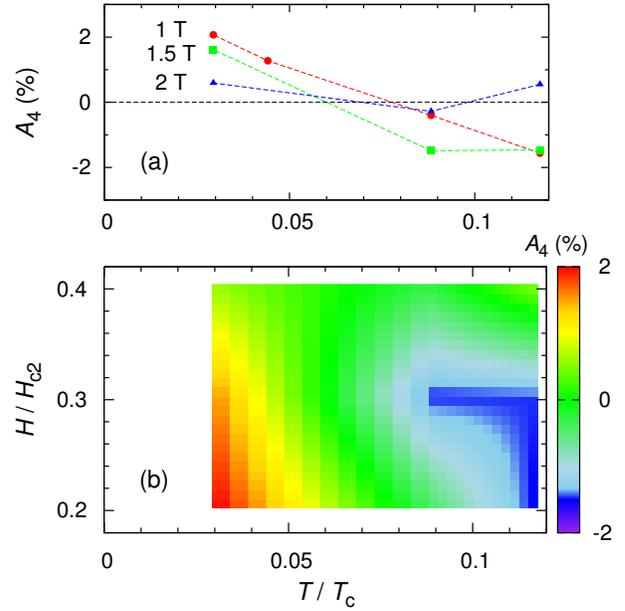}
\caption{
(Color online) 
(a) Temperature dependence of the fourfold-oscillation amplitude $A_4$ for sample C. (b) Contour plot of $A_4(T,H)$ using the data in (a).
}
\label{contour}
\end{figure}

At higher $T$, $C_{\rm e}(\phi)$ becomes maximum in $H \parallel [100]$ in contrast to the results below 0.15~K; 
a sign change of the fourfold oscillation occurs.
To characterize the $C_{\rm e}(\phi)$ oscillation, we fit $C_{\mathrm e}(\phi)$ using the function
$C_{\rm e} = C_0+C_H(1-A_4\cos4\phi)+A_2\cos2(\phi-\phi_2)$.
Here, $C_0$ and $C_H$ are the zero-field and field-dependent components of the electronic specific heat, respectively, and
$A_4$ is the amplitude of the fourfold oscillation normalized by $C_H$. 
To subtract the twofold contribution in sample~C, the function contains the term $A_2\cos2(\phi-\phi_2)$.
The fitting results are represented by dashed lines in Fig.~3 ($A_2=0$ for sample~B).
The phase shift $\phi_2$ always appears to be nearly zero.
The obtained $A_4(T,H)$ and its contour plot for sample~C are shown in Figs.~4(a) and 4(b), respectively.
In the low-field region ($\sim 0.2\Hc2$), the sign change occurs at 0.08$\Tc$.

From a theoretical viewpoint, in the vortex state, the zero-energy DOS is induced around nodes owing to the Doppler shift of $\delta E = m_{\rm e} \Vec{v}_{\rm F} \cdot \Vec{v}_{\rm s}$, 
where $m_{\rm e}$ is the electron mass and $\Vec{v}_{\rm s}$ is the local superfluid velocity that is perpendicular to $H$.\cite{Volovik1993JETPL} 
It is noted that, on the basis of the Doppler-shift analysis, \cite{Vekhter1999PRB}
when the field is applied parallel to $\Vec{v}_{\rm F}$ at a node, $\Vec{v}_{\rm F}^{\rm node}$, 
the increase in zero-energy DOS is strongly suppressed because $\delta E=0$ at the node. 
This causes local minima in $C_{\rm e}(\phi)$ when $H \parallel \Vec{v}_{\rm F}^{\rm node}$.

In addition, the microscopic theory for a nodal superconductor has predicted the occurrence of a sign change of the $C_{\rm e}(\phi)$ oscillation at about $0.1\Tc$, \cite{Vorontsov2006PRL,Hiragi2010JPSJ}
resulting from the competition between the field-angle anisotropy of zero-energy DOS and that of finite-energy DOS. 
At sufficiently low $T$, the former contribution is dominant, while the latter one, which causes a DOS oscillation of the opposite sign, becomes larger at higher $T$.
Indeed, such a sign change is observed in the $d_{x^2-y^2}$-wave superconductor CeCoIn$_5$ at $0.1\Tc$ and $0.13\Hc2$,\cite{An2010PRL} and 
the obtained $A_4(T,H)$ map of KFe$_2$As$_2$ mimics the calculated result of $A_4(T,H)$ for a $d_{xy}$-wave gap on the rippled cylindrical Fermi surface, 
corresponding to $-A_4(T,H)$ for a $d_{x^2-y^2}$-wave gap (e.g., Figs.~11-13 of Ref.~\ref{Hiragi2010JPSJ}).
Thus, the $C_{\rm e}(\phi)$ minima in $H \parallel [100]$ observed at $T \sim 0.03\Tc$ indicate 
the presence of line nodes somewhere on the Fermi surface at which $\Vec{v}_{\rm F}$ is parallel to the $[100]$ axis. 

The observed $C_{\rm e}(\phi)$ oscillation allows us to conclude that 
line nodes exist in the gap on the $\alpha$, $\zeta$, $\beta$, and/or $\epsilon$ bands at which $\Vec{v}_{\rm F} \parallel [100]$.
By the $C_{\rm e}(\phi)$ experiment alone, it is difficult to identify which of the multiple bands have nodes in the gap
because all bands have comparable masses; all the bands can contribute to $C_{\rm e}(\phi)$.
However, the present results can exclude the possibility of the $d_{x^2-y^2}$-wave symmetry,
in which line nodes are imposed in the $\Vec{k}_{\rm F} \parallel [110]$ direction 
(i.e., $\Vec{v}_{\rm F}^{\rm node} \parallel [110]$ on the $\alpha$, $\zeta$, and $\beta$ bands). 
In addition, it should be emphasized that our results are compatible with the octet-line-node scenario;~\cite{Okazaki2012Science} 
line nodes are present on the $\zeta$ band in the $\Vec{k}_{\rm F}$ direction tilted away from the $[100]$ to $[110]$ axes by $5^\circ$, 
where $\Vec{v}_{\rm F}$ is nearly parallel to the $[100]$ direction,
causing a local minimum in $C_{\rm e}(\phi)$ for $H \parallel [100]$.

In summary, the specific heat of single-crystalline KFe$_2$As$_2$ was measured down to 0.1~K in various magnetic fields and field orientations. 
Typical features of a line-node gap were observed from its field and temperature dependences.
In a rotating magnetic field within the $ab$ plane,
the low-temperature specific heat exhibited a local minimum when the field was applied along the $[100]$ direction, and
the sign of the fourfold oscillation changed at $0.08\Tc$.
On the basis of the Doppler-shift analysis, 
these results support that line nodes are located somewhere on the bands at which the Fermi velocity is oriented in the $[100]$ direction.
This is not incompatible with the octet-line-node scenario suggested from the recent laser ARPES experiments.
Moreover, an unusual upturn was found in the temperature variation of $C_{\rm e}/T$ on cooling at intermediate fields for $H \parallel ab$,
which was explained by the strong paramagnetic effect on a minor gap of the multiband superconductor.

We thank K. Okazaki and S. Shin for valuable discussions and 
T. Higo and S. Nakatsuji for supporting the X-ray diffraction analyses.
K. M. thanks F. Hardy, D. Aoki, D. A. Zocco, and K. Grube for useful discussions and information.
This work was supported by Grants-in-Aid for Scientific Research on Innovative Areas ``Heavy Electrons'' (20102007, 23102705, 21102505)
from MEXT, and KAKENHI (25800186, 21340103, 22684016, 24340090) from JSPS.

\end{document}